\documentclass[11pt,a4paper]{article}
\usepackage{jcappub}
\usepackage{rotating}
\usepackage{graphicx,epsfig}
\usepackage{amsmath}
\usepackage{xcolor}
\usepackage {amssymb}
\usepackage{subfigure}
\usepackage{time}
\usepackage{relsize}
\usepackage[utf8]{inputenc}

\title{Testing no-hair theorem by quasi-periodic oscillations: the
quadrupole of GRO J1655$-$40 }

\author[a]{ Alireza Allahyari, }
\author[b,c]{Lijing Shao}
\affiliation[a]{School of Astronomy, Institute for Research in Fundamental
	Sciences (IPM), P. O. Box 19395-5531, Tehran, Iran}
\affiliation[b]{Kavli Institute for Astronomy and Astrophysics, Peking University, Beijing 100871, China}
\affiliation[c]{National Astronomical Observatories, Chinese Academy of Sciences, Beijing 100012, China}
\emailAdd{ alireza.al@ipm.ir}
\emailAdd{lshao@pku.edu.cn}

\abstract{We perform an observational test of no-hair theorem using
quasi-periodic oscillations within the relativistic precession model. Two
well motivated metrics we apply are Kerr-Q and Hartle-Thorne metrics in
which the quadrupole is the parameter that possibly encodes deviations from
the Kerr black hole. The expressions for the quasi-periodic frequencies are
derived before comparing the models with the observation. We encounter a
degeneracy in constraining spin and quadrupole parameters that makes it
difficult to measure their values. In particular, we here propose a novel
test of no-hair theorem by adapting the Hartle-Thorne metric. We complement our study by doing a  model comparison. We find  that there is strong evidence that Kerr black hole is the source of the central object in GRO
J1655$-$40 given the present observational precisions. }

\begin{document}
\maketitle
\flushbottom

\section{Introduction}
\label{sec:intro}

Black holes (BHs) are an indispensable prediction of the theory of general
relativity (GR). Observationally they appear in various astrophysical
environments and there is wealth of direct or indirect evidence pointing
towards the existence of supermassive BHs, with masses as large as
$10^6$--$10^{10} \, M_{\odot}$. It is believed that the centers of most
sufficiently massive galaxies harbour supermassive BHs. In the current
paradigms for quasars and active galactic nuclei, as well as for the
formation of galaxies, the presence of supermassive BHs is of crucial
importance in the center of almost every galaxy~\cite{Kormendy:2013dxa}.

A generic prediction of GR is that BHs are understood to constitute the end
state of gravitational collapse of matter. In GR, a (stationary) BH final
state is given in general by the charged Kerr spacetime characterized by
its mass $M$, electric charge $e_{\rm BH}$, and angular momentum $J$. This
is known as the \textit{no-hair theorem}. According to this theorem the
multiple moments of a Kerr BH satisfy $M_{l}+{\rm i}S_{l}=M({\rm i}a)^{l}$
\cite{han74}. It is nonetheless desirable to find solutions that admit more
parameters. A test of the no-hair theorem can both identify the observed
dark compact objects with Kerr BHs and verify the validity of GR in the
strong-field regime. Given the importance of such tests, many focused
studies of the no-hair theorem have been highly regarded by many authors
\cite{Johannsen:2010bi, Tripathi:2018lhx, Psaltis:2018xkc, Liu:2018bfx,
Johannsen:2016vqy, Ni:2016rhz, Johannsen:2016uoh, Johannsen:2015mdd,
Johannsen:2013rqa, Johannsen:2012ng}.

BHs are characterized by the existence of the event horizon, the boundary
beyond which even light is unable to escape. The existence of the event
horizon has the advantage that the exterior region satisfies the principle
of causality. To preserve this, Penrose has conjectured that the
singularities that would actually occur in nature must be BHs. This is
known as the Cosmic Censorship Conjecture. However, there is no general
proof for this conjecture and more general collapsed configurations are
possible within the classical framework of GR.

The other evidence in favor of existence of BHs is the detection of
gravitational waves from astrophysical sources. Most detections of
gravitational waves have been interpreted in terms of binary BH mergers
\cite{Abbott:2016blz, Abbott:2016nmj, TheLIGOScientific:2016pea,
Abbott:2017gyy, Abbott:2017vtc, Abbott:2017oio}.

Evidently, extremely compact gravitationally collapsed systems definitely
exist, but the observational evidence regarding their precise nature is
somehow inconclusive. We therefore turn to compare the theoretical evidence
based on Einstein's GR with observations.

One yet promising arena to test the nature of the compact objects is by
quasi-periodic oscillations (QPOs) observed in the X-ray flux emitted by
accreting BHs in X-ray binary systems \cite{Abramowicz:2001bi,
Pasham:2015tca}. In these systems, a BH or a neutron star (NS) accretes
materials from a stellar companion \cite{Pasham:2018bkt}. QPOs are observed
as narrow features in the power spectra of the light curves of accreting
NSs and BHs \cite{Belloni:2012sv}. The very first hint towards their
discovery in the literature dates back to the results reported in
ref.~\cite{samimi} and a definite discovery in ref.~\cite{rebu}. There are
various mechanisms proposed to account for such narrow features in the
power spectra ranging from relativistic precession models, diskoseismology
models, resonance models and $p$-mode oscillations of an accretion torus
\cite{Stella:1997tc, Stella:1998mq, Stella:1999sj, Perez:1996ti,
Silbergleit:2000ck, Ingram:2020aki, Motta:2017sss, Bambi:2016iip,
Stuchlik:2008fy, Rezzolla:2003zx}. These attempts have tried to pin down
the mechanism that could account for the QPO features.

One of the highly regarded models for QPOs is the relativistic procession
model according to which the QPO frequencies are believed to be related to
fundamental frequencies of test particles orbiting a central object
\cite{Stella:1997tc,Stella:1999sj}. These particles undergo small
oscillations when perturbed as they revolve around the central object. This
model was later extended in ref.~\cite{miller}. The promising aspect of
using the QPOs as a method to study the nature of compact objects is that
they can be measured with rather high precision \cite{Maselli:2017kic,
Franchini:2016yvq, Suvorov:2015yfv, Boshkayev:2015mna, Cardoso:2019rvt,
Maselli:2014fca}. The crucial fact about QPOs is that they can be used to
probe the strong filed regime of GR \cite{Berti:2015itd, Psaltis:2008bb,
Psaltis:2009xf, Jusufi:2020odz, Azreg-Ainou:2020bfl}.

Although it appears that modeling an accreting object with a Kerr metric is
a good approximation, the exact nature of the compact object deserves a
thorough investigation. We seek to model an accreting object associated
with the QPOs as a rotating metric with a small quadrupole. Our models are
based on the Kerr-Q metric and Hartle-Thorne metric~
\cite{HaTh,Toktarbay:2015lua,Allahyari:2019umx}. They represent a slowly
rotating object with the quadrupole $q$ and the physical mass $M$ and the
rotation $a$. In a previous paper, Allahyari {\it et al.} have investigated
the properties of Kerr-Q metric and derived the quasi normal frequencies of
this metric \cite{Allahyari:2018cmg, Allahyari:2019umx}. Here, we wish to
constrain the quadrupole of the central object in GRO J1655$-$40 by its
observed QPOs~\cite{Motta:2013wga}. There are various attempts to generalize the Kerr metric. Let us highlight the importance of our metrics and mention their motivations. The reader could see refs. \cite{Allahyari:2018cmg, Allahyari:2019umx} for more detailed discussions.  The Kerr-Q is derived from an exact solution of vacuum Einstein equations. It has been tested to be singularity free outside the horizon and is asymptotically flat. It is also the simplest extension of the Kerr metric. In this way, it has desirable properties and is originated from GR rather than some  phenomenological parameterization.
The Hartle-Thorne metric is on the other hand an approximate solution of GR equations  which could originate from an inner source. The quadrupole of the Hartle-Thorne is induced by the inner source configuration. It is crucial to mention why we use two classes of metrics. We use these two metrics to model two separate situations. The Hartle-Thorne metric could describe an inner source for the compact object, whereas  the Kerr-Q metric is a  generalized BH. There are various vacuum metrics with higher order moments. A review of the exact metrics with motivated properties can be found in ref.~\cite{Frutos-Alfaro:2017ops}. It has been shown that they are  equivalent up to the level of the quadrupole moment. More specifically, a connection between  Quevedo-Mashhoon metric \cite{MQ}  and the Hartle-Thorne metric has been established in ref.~\cite{Quevedo:2010vx}. Gutsunaev and Manko \cite{GM} found a metric which was similar to the Erez-Rosen metric \cite{ER} up to the quadrupole moment in ref.~\cite{Q}. The other metric is the metric in ref.~\cite{Johannsen:2011dh}. This metric is derived by applying a Newman-Janis algorithm to a seed metric. The resulting metric is not a vacuum solution in GR which is our case study here. A metric similar to the Hartle-throne metric we have used here is presented in ref.~\cite{Glampedakis:2005cf} where they bring the Hartle-Throne metric to the Boyer-Lindquist coordinates.

This manuscript is structured as follows. In section.~\ref{sec:background} we
give a general review of rotating $\delta$-metric and expand it to first
order in the quadrupole parameter and define QPOs for this metric. In
section.~\ref{HTsec}, we review the Hartle-Thorne metric and its QPOs. In
section.~\ref{constaint-sec},  we constrain the quadrupole in our models
by comparing with the observed QPOs for GRO J1655$-$40. In section.~\ref{m-c}, we compare the models with the Kerr metric using the Bayes factors. Section.~\ref{s-d} is devoted to the summary and discussion.  Across this
manuscript, for the signature of the spacetime metric, we will implement
the $(-, +, +, +)$ convention.

\section{Quadrupolar metrics}

In this section we introduce two metrics that admit a quadruple, namely the
Kerr-Q metric and the Hartle-Thorne metric. They both describe a
configuration to first order in the quadrupole and to second order in the
rotation. $q$-metric is derived from the rotating $\delta$-metric by writing $\delta=1+q$ and $m=M/(1+q)$ and expanding to first order in $q$ where $M$ is the physical mass as we will explain. If we repeat the same steps for rotating $\delta$-metric and additionally expand to second order in the rotation parameter $a$, we get Kerr-Q metric ~\cite{Allahyari:2018cmg, Allahyari:2019umx}.
These metrics were derived from the exact solutions by expanding to the desired order. There is another solution with a quadrupole which is not derived from an exact solution. This metric is the Hartle-Thorne solution. 

\subsection{Rotating $\delta$-metric}\label{sec:background}

We start with the $\delta$-metric and explain how to derive the Kerr-Q metric.
The rotating
$\delta$-metric is a generalized Kerr metric that admits a quadrupole.  This
metric is obtained by rotating the $\delta$-metric as the seed metric using
solution generating techniques involving the Hoenselaers-Kinnersley-Xanthopoulos (HKX) 
transformations in which symmetries of the field equations are used to find solutions  that admit  rotation, multipole moments and  magnetic dipole from given seed metrics~\cite{ Frutos-Alfaro:2016arb, Toktarbay:2015lua, Hoenselaers:1979mk}. 

Supposing that the central object is axisymmetric, we focus on the metrics with this symmetry. 
The general stationary axisymmetric  line element can be represented in spheroidal coordinates  $(t, x, y, \phi)$ as
\begin{eqnarray}
\label{stationary}
ds^2 & = & - f (d t - \omega d \phi)^2 \\
& + & \frac{\sigma^2}{f}
\left[{\rm e}^{2 \gamma} (x^2 - y^2) \left(\frac{d x^2}{x^2 - 1}
+ \frac{d y^2}{1 - y^2} \right) + (x^2 - 1)(1 - y^2) d \phi^2 \right] ,
\nonumber
\end{eqnarray}
where  $ \sigma $ is a positive constant and all the metric functions, $f$, $\omega$ and $\gamma$, depend on $x$ and $y$. In spheroidal coordinates they are
\begin{eqnarray}
\label{q0}
f & = & \frac{{\cal A}}{{\cal B}} , \nonumber \\
\omega & = & - 2 \left(-a + \sigma \frac{{\cal C}}{{\cal A}} \right),  \\
{\rm e}^{2 \gamma} & = & \frac{1}{4} \left(1 + \frac{m}{\sigma} \right)^{2}
\frac{{\cal A}}{(x^{2} - 1)^{1 + q}} 
{\left[\frac{x^{2} - 1}{x^{2} - y^{2}} \right]}^{(1 + q)^{2}} , \nonumber 
\end{eqnarray}
where we have defined
\begin{eqnarray}
\label{q1}
{\cal A} & = & a_{+} a_{-} + b_{+} b_{-} , \nonumber \\ 
{\cal B} & = & a_{+}^{2} + b_{+}^{2} , \\
{\cal C} & = & \! (x + 1)^{q} \left[x (1 - y^{2})(\lambda + \eta) a_{+} 
+ y (x^{2} - 1)(1 - \lambda \eta) b_{+} \right] , \nonumber \\ 
\end{eqnarray}
along with 
\begin{eqnarray}
\label{q2}
a_{\pm} & = & (x \pm 1)^{q} 
[x (1 - \lambda \eta) \pm (1 + \lambda \eta)] , \nonumber \\
b_{\pm} & = & (x \pm 1)^{q} [y (\lambda + \eta) \mp (\lambda - \eta)] , 
\end{eqnarray}
and
\begin{eqnarray}
\label{q3}
\lambda & = & \alpha (x^{2} - 1)^{- q} (x + y)^{2 q} ,  \\
\eta & = & \alpha (x^{2} - 1)^{- q} (x - y)^{2 q} , \nonumber
\end{eqnarray}
with
$\;\alpha a  = {m-\sigma} $.

We recover the $q$-metric  when $a$ tends to zero
\cite{Allahyari:2018cmg,Toshmatov:2019qih}. We have also checked that when
$q=0$, we recover the Kerr metric in Boyer-Lindquist coordinates by using
the following transformation
\begin{align}
x&=\frac{r-m}{\sigma},\\
y&=\cos \theta,\nonumber \\ 
\sigma&=\sqrt{m^2-a^2}.\nonumber
\end{align}
Finally we transform the metric to spherical coordinates. The metric in
spherical coordinates can be written as 
\begin{align}
ds^2&=-fdt^2+2f\omega dtd\phi+\frac{{\rm e}^{2 \gamma}}{f}\frac{\mathbb{B}}{\mathbb{A}} dr^2+r^2\frac{{\rm e}^{2 \gamma}\;\mathbb{B}}{f}d\theta^2\\\nonumber
&+\left[\frac{r^2\mathbb{A}\sin^2\theta }{f}-f\omega^2 \right]d\phi^2, 
\end{align}
where we have
\begin{align}
\mathbb{A}&=1-\frac{2 m}{r}+\frac{a^2}{r^2},\\
\mathbb{B}&=1-\frac{2m}{r}+\frac{a^2}{r^2}+\frac{\sigma^2\sin^2\theta}{r^2}.\\\nonumber
\end{align}
Note that 
\begin{align}
\frac{r^2}{\sigma^2}\mathbb{A}&=-1+\left(\frac{r-m}{\sigma} \right)^2 .
\end{align}

\subsubsection{Mass and quadrupole }

For small  values of $a$ and $q$  we may expand the metric to first order in $q$ and to second order in $a$. We also neglect terms like $q \,a^2$ and $q\, a$. We have
\begin{align}\label{met1}
g_{tt}&=-\hat{\mathbb{A}}\left(1+q\ln \hat{\mathbb{A}} \right)-\frac{2 a^2 m \cos^2\theta}{r^3},\\
g_{t\phi}&=-\frac{2am\sin^2\theta}{r},\\
g_{rr}&=\dfrac{1}{\hat{\mathbb{A}}}\left(1+q\ln\frac{\hat{\mathbb{A}}}{\hat{\mathbb{B}}^2} \right) -a^2\frac{1-\cos^2\theta\hat{\mathbb{A}}}{r^2\hat{\mathbb{A}}^2},\\
g_{\theta\theta}&=r^2+a^2\cos^2\theta+qr^2\ln\frac{\hat{\mathbb{A}}}{\hat{\;\mathbb{B}}^2},
\end{align}
\begin{align}\label{met2}
g_{\phi\phi}&=\left[ r^2-qr^2\ln\hat{\mathbb{A}}+a^2\left( 1+\frac{2m\sin^2\theta}{r}\right) \right]  \sin^2\theta,
\end{align}
where 
\begin{align}
\hat{\mathbb{A}}&:=1-\frac{2m}{r},\\
\hat{\;\mathbb{B}}&:=1-\frac{2m}{r}+\frac{m^2\sin^2\theta}{r^2}.
\end{align}
This metric represents the superposition of Kerr metric and $q$-metric.
When $a=0$, we get the $\delta$-metric  expanded to first order
in $q$~\cite{Allahyari:2018cmg}. When $q=0$, we get the Kerr metric to
second order in $a$. The mass and quadrupole of this metric have been found
after taking the weak field limit and a coordinate transformation by which
the metric takes the Newtonian form. Therefore, we have for the physical
mass, quadrupole, and angular momentum,
\begin{align}
\label{def}
M=m(1+q),\qquad Q=\frac{2}{3}M^3 q+a^2M,\qquad J=Ma.
\end{align} 
Therefore, the physical mass $M$ is different from $m$. Also note that the
parameter $q$ gives the quadrupole of the metric \cite{ Allahyari:2019umx}.
Following the discussion in ref.~\cite{Allahyari:2018cmg}, for oblate
(prolate) configurations we have $q>0$ ($q<0$). Also note that we expect
the oblateness for a rotating mass $q>0$. Hence, we will assume $q>0$.

\subsubsection{Kerr-$Q$ metric}\label{kerrq}

Note that in the rotating $\delta$-metric in eqs.~\eqref{met1}--\eqref{met2}, the mass
parameter $m$ is not physical. We wish to adapt it to a more desirable form
expanded in terms of the physical mass $M$. To cast the metric in terms of
the physical mass, we must take the  rotating $\delta$-metric 
and replace $m$ by $M/\left( 1+q\right) $. After expanding the resulting
metric to linear order in $q$ and to second order in $a$, we obtain
\begin{align}\label{R6}
ds_{\rm KQ}^2 ={}& -  \left[\tilde{\mathbb{A}}+q\left(\frac{2M}{r\tilde{\mathbb{A}}} +\ln{\tilde{\mathbb{A}}}\right)\tilde{\mathbb{A}}+\frac{2 a^2 M \cos^2\theta}{r^3}\right]\,dt^2-\frac{4a M\sin^2\theta}{r}dt\,d\phi\\
&+ \left[\frac{1}{\tilde{\mathbb{A}}}-\frac{q}{\tilde{\mathbb{A}}}\left(\frac{2M}{r\tilde{\mathbb{A}}} +\ln{\frac{\tilde{\mathbb{B}}^2}{\tilde{\mathbb{A}}}}\right)-a^2\frac{1-\cos^2\theta\tilde{\mathbb{A}}}{r^2{\tilde{\mathbb{A}}}^2}\right]\,dr^2 \nonumber \\
\nonumber & + \left(1-q\ln{\frac{\tilde{\mathbb{B}}^2}{\tilde{\mathbb{A}}}}+\frac{a^2}{r^2}\cos^2\theta\right)\,r^2\,d\theta^2 + \left[  1-q\ln{\tilde{\mathbb{A}}}+\frac{a^2}{r^2}\left( 1+\frac{2M\sin^2\theta}{r}\right) \right]  \,r^2\sin^2\theta\,d\phi^2\,,
\end{align}
where
\begin{align}
\tilde{\mathbb{A}}&:=1-\frac{2M}{r},\\
\tilde{\;\mathbb{B}}&:=1-\frac{2M}{r}+\frac{M^2\sin^2\theta}{r^2}.
\end{align}
This metric reduces to the Kerr metric (to second order in $a$) when $q=0$.
Let us emphasize that this metric has different components from the
rotating $\delta$-metric in eqs.~\eqref{met1}--\eqref{met2}, as we have
rewritten it in terms of the physical mass $M$ rather than the non-physical
mass parameter $m$. We will implement this metric in our calculations.

\subsubsection{QPOs for Kerr-$Q$ metric}

\begin{figure}[!ht]
	\begin{center}
		\includegraphics[width=0.3\linewidth]{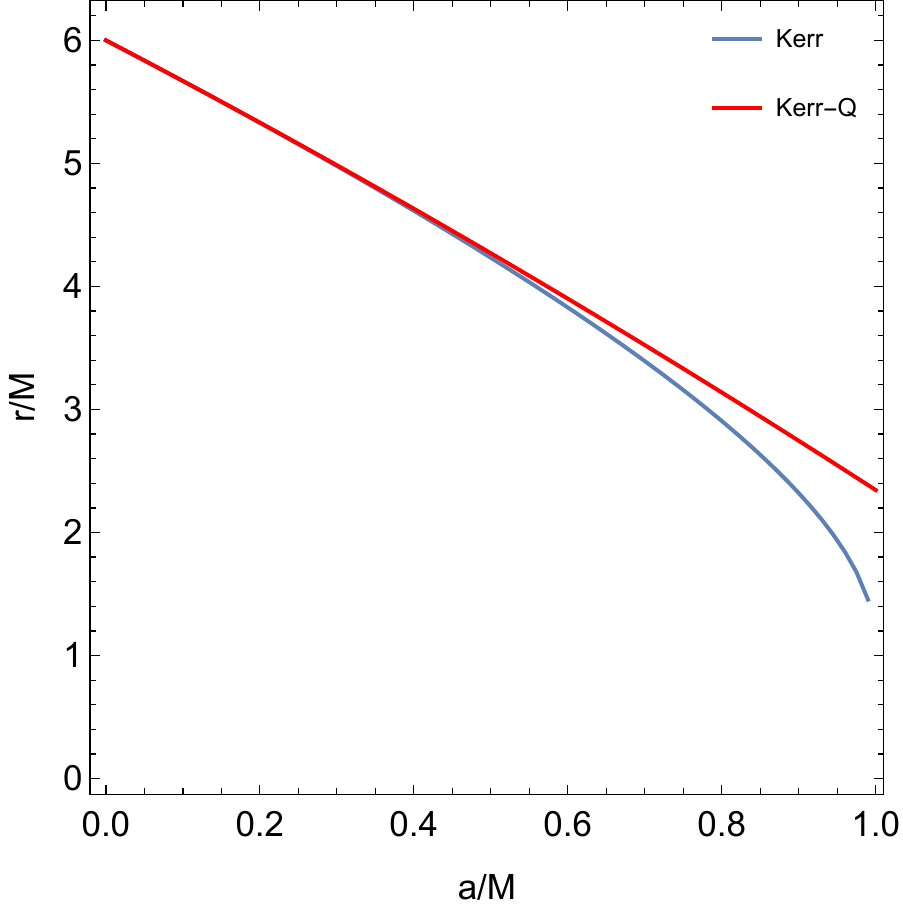}
		\caption{ $r_{\rm ISCO}$  for Kerr and Kerr-$Q$ metrics. }
		\label{kerr}
	\end{center}
\end{figure}

Let us consider a test particle in orbit around a stationary axisymmetric
source. The general line element in spherical coordinates is given by
\begin{equation}
ds^2 = g_{tt} dt^2 + g_{rr}dr^2 + g_{\theta\theta} d\theta^2 
+ 2g_{t\phi}dt d\phi + g_{\phi\phi}d\phi^2 \ ,
\end{equation}
where all functions depend on $r$ and $\theta$.
There are two constants of motion imposed by the symmetries of the metric, the specific energy at infinity, $E$, and the $z$-component of specific angular momentum $L_z$. Therefore, we have 
\begin{eqnarray}
\dot{t} &= \frac{E g_{\phi\phi} + L_z g_{t\phi}}{
	g_{t\phi}^2 - g_{tt} g_{\phi\phi}} \,\label{eqt} , \quad \\
\dot{\phi}& = - \frac{E g_{t\phi} + L_z g_{tt}}{
	g_{t\phi}^2 - g_{tt} g_{\phi\phi}} \,\label{eq2} \; ,
\end{eqnarray}
where dot represents the derivative with respect to the affine parameter.
For a timelike path, $g_{\mu\nu}\dot{x}^\mu \dot{x}^\nu = -1$. After using
eqs.~\eqref{eqt} and ~\eqref{eq2}, we have
\begin{eqnarray}
g_{rr}\dot{r}^2 + g_{\theta\theta}\dot{\theta}^2
= V_{\rm eff}(r,\theta,E,L_z)\label{vpot} \, ,
\end{eqnarray}
where the effective potential $V_{\rm eff}$ is
\begin{eqnarray}
V_{\rm eff} = \frac{E^2 g_{\phi\phi} + 2 E L_z g_{t\phi} + L^2_z 
	g_{tt}}{g_{t\phi}^2 - g_{tt} g_{\phi\phi}} - 1  \, .
\end{eqnarray}
We are interested in the equatorial circular orbits with $\theta=\pi/2$ and $\dot{r}=\ddot{r}=\dot{\theta}=0$. The radial component of the geodesic equations gives
\begin{eqnarray}
\Omega_\phi =\frac{\dot{\phi}}{\dot{t}} =\frac{- \partial_r g_{t\phi} 
	\pm \sqrt{\left(\partial_r g_{t\phi}\right)^2 
		- \left(\partial_r g_{tt}\right) \left(\partial_r 
		g_{\phi\phi}\right)}}{\partial_r g_{\phi\phi}} \, ,
\end{eqnarray}
where $+/ -$ corresponds to co-rotating/counter-rotating orbits, respectively.
The orbital frequency is thus $\nu_\phi= \Omega_\phi/2\pi$.
For the especial case of the circular orbits, we have the expressions
\begin{eqnarray}
E &=& - \frac{g_{tt} + g_{t\phi}\Omega_\phi}{
	\sqrt{-g_{tt} - 2g_{t\phi}\Omega_\phi - g_{\phi\phi}\Omega^2_\phi}} \, , \\
L_z &=& \frac{g_{t\phi} + g_{\phi\phi}\Omega_\phi}{
	\sqrt{-g_{tt} - 2g_{t\phi}\Omega_\phi - g_{\phi\phi}\Omega^2_\phi}} \, ,
\end{eqnarray}
for a particle orbiting at the constant radius $r_0$. One important
parameter to take into account is the innermost stable circular orbit
(ISCO). In the Schwarzschild metric this is simply $r_{\rm ISCO}=6 M$. The
position of ISCO for Kerr-$Q$ metric can be found perturbatively by using
$\frac{d^2 V_{\rm eff}}{dr^2}=0$. After some algebra, we obtain
\begin{align}
r_{\rm ISCO}=-\frac{7 a^2}{18 M}-4 \sqrt{\frac{2}{3}} a-\frac{M q}{2}+6 M.
\end{align}
This expression reduces to $r_{\rm ISCO}=6 M$ for the Schwarzschild metric
where $a=q=0$. In figure.~\ref{kerr} we have plotted the $ r_{\rm ISCO}/M  $ as a function of $a/M$ for the Kerr metric and the Kerr-$Q$ metric for $q=0$. This figure shows the validity of the expansion to $(a/M)^2$ in our calculations. 

\begin{figure}[!ht]
	\begin{center}
		\includegraphics[width=0.75\linewidth]{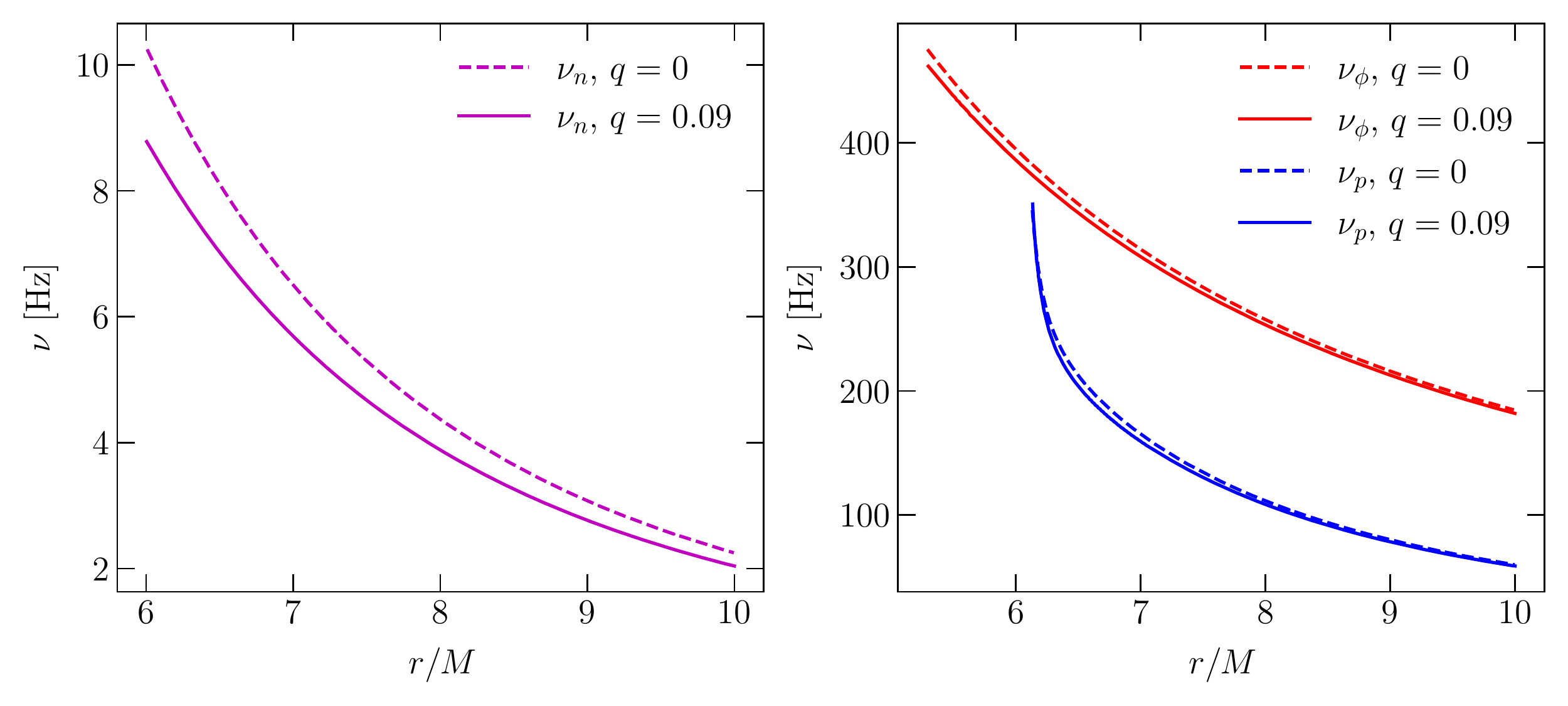}
		\caption{\textit{Left}: nodal frequency as a function of $r/M$.
		\textit{Right}: $\nu_{\phi}$ (red) and $\nu_p$ (blue) as a function
		of $r/M$. We have assumed $M/M_\odot=5.5$ and $a/M$=0.2.}
		\label{SH1}
	\end{center}
\end{figure}

\begin{figure}[!ht]
		\hspace{-1.1cm}\includegraphics[width=1.1\linewidth]{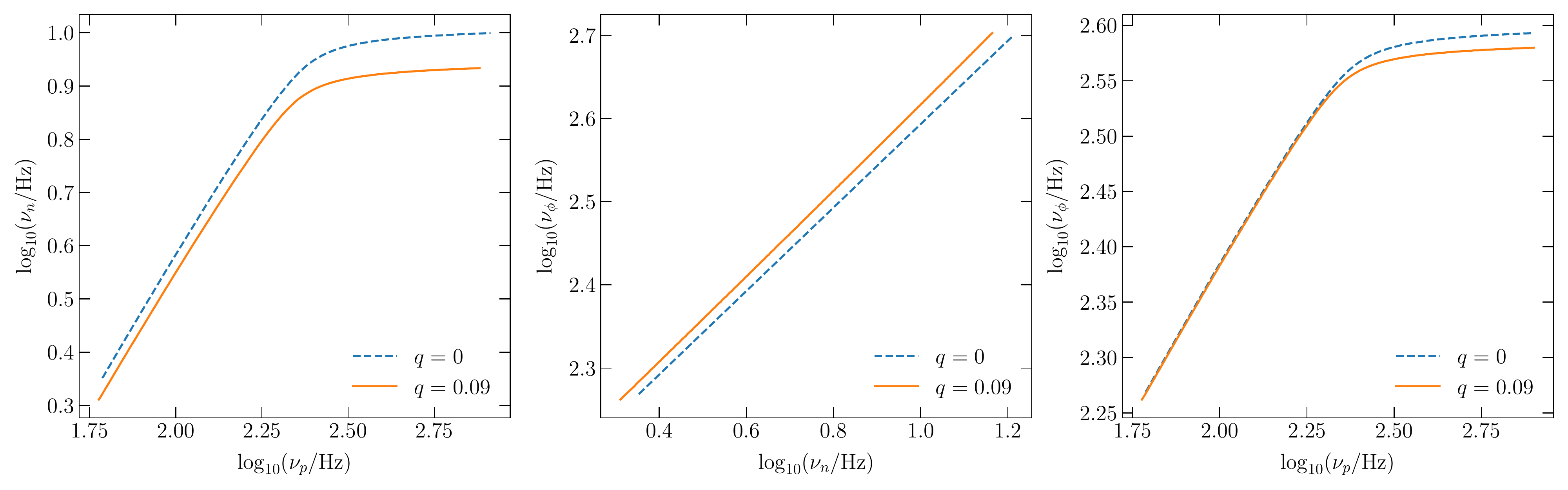}	
		\caption{Interdependence of three frequencies. }
		\label{SH1-1}
\end{figure}

 Let us suppose that the particle orbit is slightly perturbed
along $r$ and $\theta$ directions. We have $r=r_0\left(1+\delta_r \right) $
and $\theta=\pi/2+\delta_{\theta}$. Using eq.~\eqref{vpot}, we find that
the equations for perturbations are given by
\begin{eqnarray}
\label{eq-o1}
\frac{d^2 \delta_r}{dt^2} + \Omega_r^2 \delta_r = 0 \, , \quad
\frac{d^2 \delta_\theta}{dt^2} + \Omega_\theta^2 \delta_\theta = 0 \, ,
\label{eq-o2}
\end{eqnarray}
where the frequencies of the oscillations are
\cite{Merloni:1998by,Maselli:2017kic}
\begin{eqnarray}
\Omega^2_r = - \frac{1}{2 g_{rr} \dot{t}^2} 
\frac{\partial^2 V_{\rm eff}}{\partial r^2} \, , \quad
\Omega^2_\theta = - \frac{1}{2 g_{\theta\theta} \dot{t}^2} 
\frac{\partial^2 V_{\rm eff}}{\partial \theta^2} \, .
\label{eq-ot}
\end{eqnarray}
According to the precession model~\cite{Stella:1997tc}, the QPOs are
related to the periastron precession frequency and nodal precession
frequency. The periastron precession frequency $\nu_{\rm p}$ and the nodal
precession frequency $\nu_{\rm n}$ can be obtained as
\begin{eqnarray}
\nu_{\rm p} = \nu_\phi - \nu_r \, , \quad
\nu_{\rm n} = \nu_\phi - \nu_\theta \, .
\end{eqnarray}
figures~\ref{SH1} and~\ref{SH1-1} show the orbital, periastron and nodal
frequencies for the Kerr-$Q$ metric where we have used $M/M_\odot=5.5$ and
$a/M$=0.2. We have shown the $q=0$ case for comparison. Figure~\ref{SH1-1}
shows the relations of QPO frequencies. It is important to note that as we move away from the source the frequencies fall off, and because the quadrupole term falls as $1/r^3$ , the frequencies approach the Kerr frequencies.


\subsection{Rotating Hartle-Thorne metric} \label{HTsec}

The other solution that treats quadrupole to linear order and rotation to
second order is the Hartle-Thorne metric~\cite{HaTh}. This metric is given
by
\begin{equation}\label{HT}
ds_{\rm RHT}^2 = -\mathbb{F}_1 \,dt^2 + \frac{1}{\mathbb{F}_2}\,dr^2 +  \mathbb{G}\,r^2\,\left[d\theta^2 +\sin^2\theta\,\left(d\phi - \frac{2J}{r^3}\,dt\right)^2\right]\,,
\end{equation}
where
\begin{equation}\label{L2}
\mathbb{F}_1 = \left(1 -\frac{2M}{r}+ \frac{2J^2}{r^4}\right)\,\left[ 1 +  \frac{2J^2}{Mr^3}\left(1+\frac{M}{r}\right)P_2(y)+ 2\tilde{q}\, \mathcal{Q}_2^2(x)\,P_2(y)\right]\,,
\end{equation}
\begin{equation}\label{L3}
\mathbb{F}_2 = \left(1 -\frac{2M}{r}+ \frac{2J^2}{r^4}\right)\,\left[ 1 +  \frac{2J^2}{Mr^3}\left(1-5\,\frac{M}{r}\right)P_2(y)+ 2\tilde{q}\, \mathcal{Q}_2^2(x)\,P_2(y)\right]\,,
\end{equation}
\begin{equation}\label{L4}
\mathbb{G} = 1 -  \frac{2J^2}{Mr^3}\left(1+2\,\frac{M}{r}\right)P_2(y)+ 2\tilde{q}\,\left[\frac{2M}{\sqrt{r(r-2M)}}\,\mathcal{Q}_2^1(x)-\mathcal{Q}_2^2(x)\right]\,P_2(y)\,,
\end{equation}
where  $x = -1+ r/M$, $y = \cos\theta$, and $\tilde{q}$ is defined by 
\begin{equation}\label{L5}
\tilde{q} := \frac{5}{8}\,\frac{Q-J^2/M}{M^3}=q-\frac{5J^2/M}{8M^3}\,.
\end{equation}
The definitions for the Legendre functions of the second kind in interval
$x \in [1, \infty)$ are given by
\begin{equation}\label{H5}
\mathcal{Q}_2^1(x) =  - (x^2-1)^{1/2}\,\left[\frac{3}{2} \,x\,  \ln\left(\frac{x+1}{x-1}\right) -\frac{3x^2-2}{x^2-1}\right]\,,
\end{equation}
\begin{equation}\label{H6}
\mathcal{Q}_2^2(x) =  \frac{3}{2}\, (x^2-1)\,\ln\left(\frac{x+1}{x-1}\right) -x\,\frac{3x^2-5}{x^2-1}\,.
\end{equation}
It is shown that when the quadrupole moment is given by $Q_K = J^2/M$, that
is $q_K=\frac{5a^2}{8M^2}$, the metric will reduce to the Kerr metric in
Boyer-Lindquist coordinates after a coordinate transformation
\cite{Allahyari:2019umx}.

\begin{figure}[!ht]
	\begin{center}
		\includegraphics[width=0.75\linewidth]{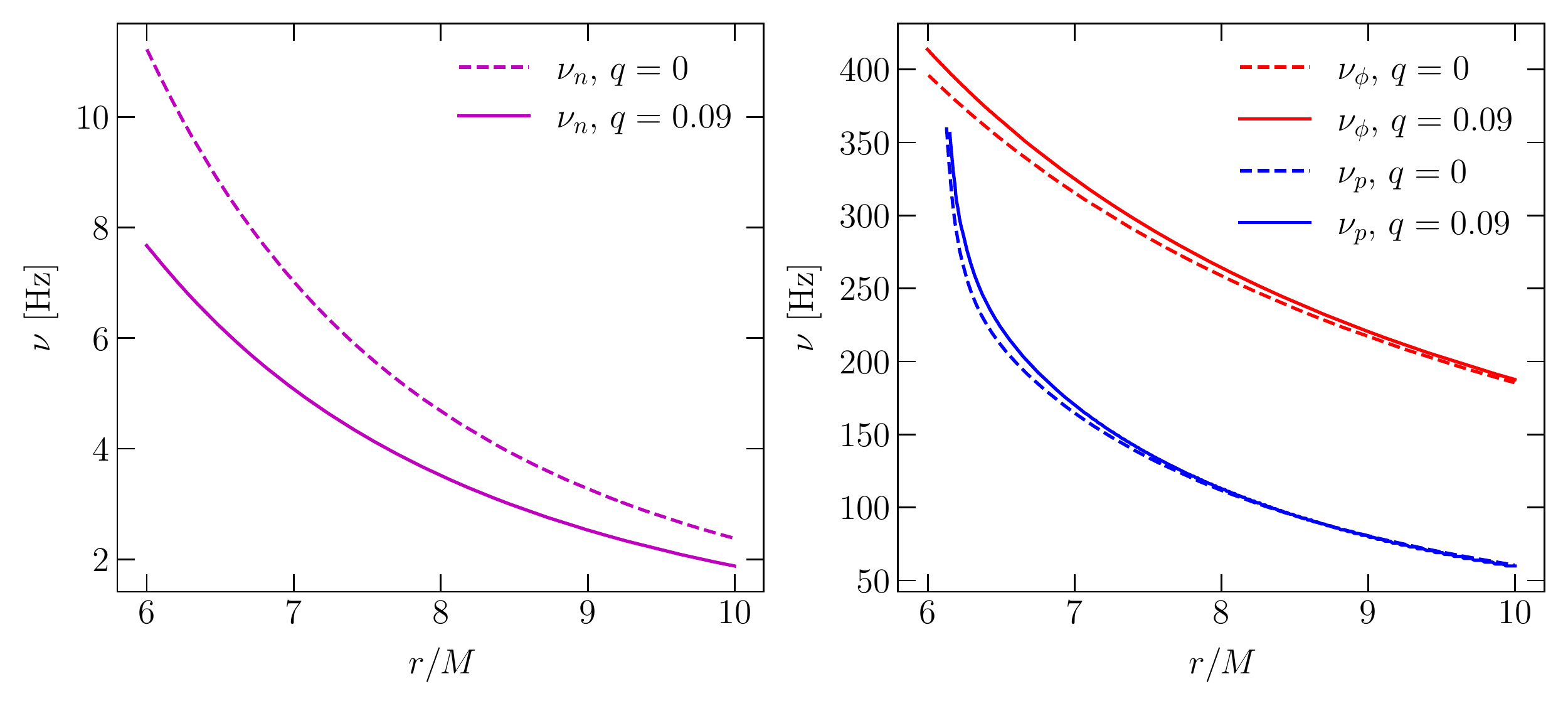}
		\caption{\textit{Left}: nodal frequency as a function of $r/M$ for
		the Hartle-Thorne metric. \textit{Right}: $\nu_{\phi}$ (red) and
		$\nu_p$ (blue) as a function of $r/M$. We have assumed
		$M/M_\odot=5.5$ and $a/M$=0.2.}
		\label{SH3}
	\end{center}
\end{figure}

\begin{figure}[!ht]
		\hspace{-1.1cm}\includegraphics[width=1.1\linewidth]{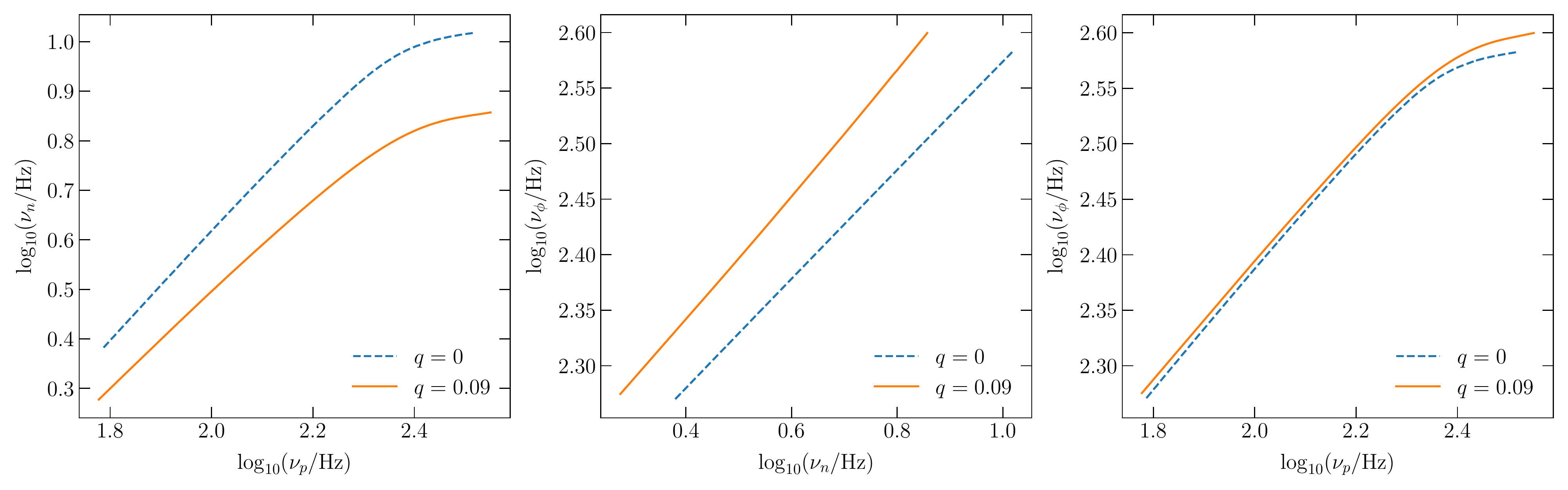}	
		\caption{Interdependence of three frequencies for the Hartle-Thorne
		metric.}
		\label{SH3-1}
\end{figure}

\subsubsection{QPOs for rotating Hartle-Thorne metric}

Figure~\ref{SH3} shows the QPOs for the Hartle-Thorne metric where we have
used $M/M_\odot=5.5$ and $a/M$=0.2. We have shown the $q_K$ case for
comparison with the Kerr metric. Again we see that frequencies fall off and the frequencies approach the Kerr values as we move away from the center. Figure~\ref{SH3-1} shows the
interdependence of three frequencies, as in figure.~\ref{SH1-1}. In a similar
way to the Kerr-$Q$ metric, the ISCO for the Hartle-Thorne can be derived.
The expression for the ISCO reads 
\begin{align}
r_{\rm ISCO}=-\frac{1.35355 a^2}{M}-4 \sqrt{\frac{2}{3}} a+\frac{1.05726 Q}{M^2}+6 M\,.
\end{align}

\section{Constraints by QPOs }\label{constaint-sec}

In this section our aim is to fit the metrics proposed in the previous
sections to investigate the quadrupole of GRO~J1655$-$40. GRO~J1655$-$40 is
an X-ray binary system where one of the stars is probably a BH
\cite{Orosz:1996cg}. The measurement of QPOs for this source has been
realized through RXTE observations \cite{Strohmayer, Remillard:1998ah}.
Three QPO frequencies have been measured using X-ray timing method
\cite{Motta:2013wga, Ingram:2020aki}. We assume that this compact object is
not a Kerr BH but a rotating source with a quadruple described by the
Kerr-$Q$ metric or the Hartle-Thorne metric. To find the best parameters we
use a Bayesian approach. Assuming that the likelihood is given by $\cal
L\sim\rm e^{-\chi^2/2}$, we use the $\chi$-square defined in
ref.~\cite{Bambi:2013fea}. The $\chi$-square takes the following form
\begin{eqnarray}
\chi^{2}(a,q,M,r)=& 
\frac{\left( \nu_{\rm C} - \nu_{\rm n} \right)^2}{\sigma^2_{\rm C}}
+ \frac{\left( \nu_{\rm L} - \nu_{\rm p} \right)^2}{\sigma^2_{\rm L}}
+ \frac{\left( \nu_{\rm U} - \nu_\phi \right)^2}{\sigma^2_{\rm U}} \, ,
\label{eq-chi20}
\end{eqnarray}
where $\nu_{\rm C}$, $\nu_{\rm L}$ and $\nu_{\rm U}$, as well as their errors denoted
by $\sigma^2_{i}$ ($i \in \{\rm C, L, U\}$), are provided by the
observations. Let us mention that we should not compare the values of $r$
in different metrics we present here, as they have different meaning
associated with the underlying geometry and are not physical observables
unlike the quadrupoles which are measurable by inertial observers at
infinity.

For GRO~J1655$-$40 the corresponding frequencies have been measured based
on RossiXTE observations~\cite{Motta:2013wga} as

\begin{eqnarray}
\nu_{\rm C} = 17.3 \; {\rm Hz} \, , && \sigma_{C} = 0.1 \; {\rm Hz} \, , \\
\nu_{\rm L} = 298 \; {\rm Hz} \, , && \sigma_{L} = 4 \; {\rm Hz} \, , \\
\nu_{\rm U} = 441 \; {\rm Hz} \, , && \sigma_{U} = 2 \; {\rm Hz} \, .
\label{tab1}
\end{eqnarray}
\begin{figure}[!ht]
	\begin{center}	
		\includegraphics[width=0.8\linewidth]{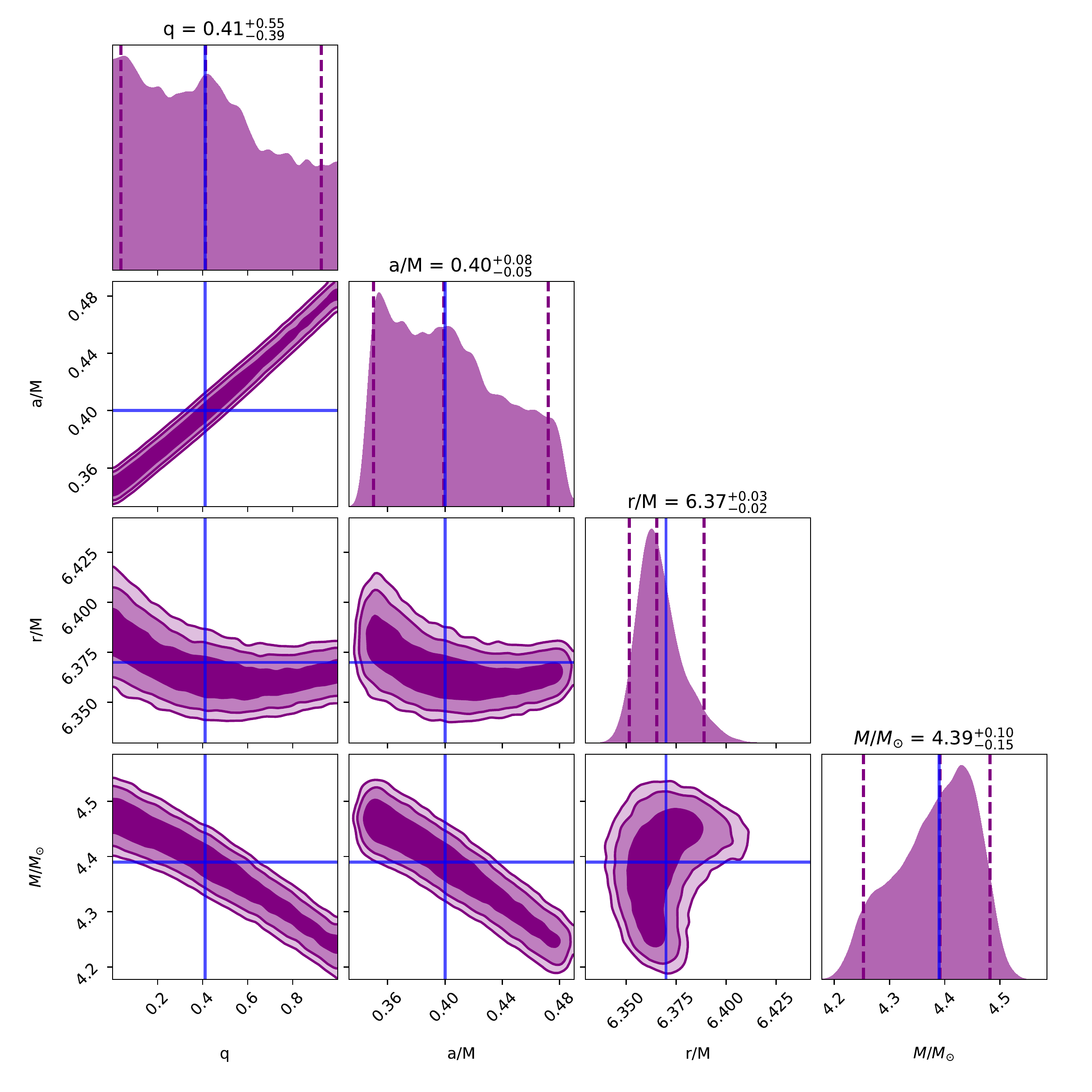}
		\caption{Two dimensional marginal posteriors for parameters in the
		Kerr-$Q$ metric. The one dimensional distributions are also shown
		by the diagonal figures. The contours show 68\%, 95\% and 99\%
		credible intervals. Solid lines represent the mean values. Dashed lines are the
		5\%, 50\%, and 95\% percentiles of the distribution. }
		\label{figmc-2}		
	\end{center}
\end{figure}

\begin{figure}[!ht]
	\begin{center}		
		\includegraphics[width=0.8\linewidth]{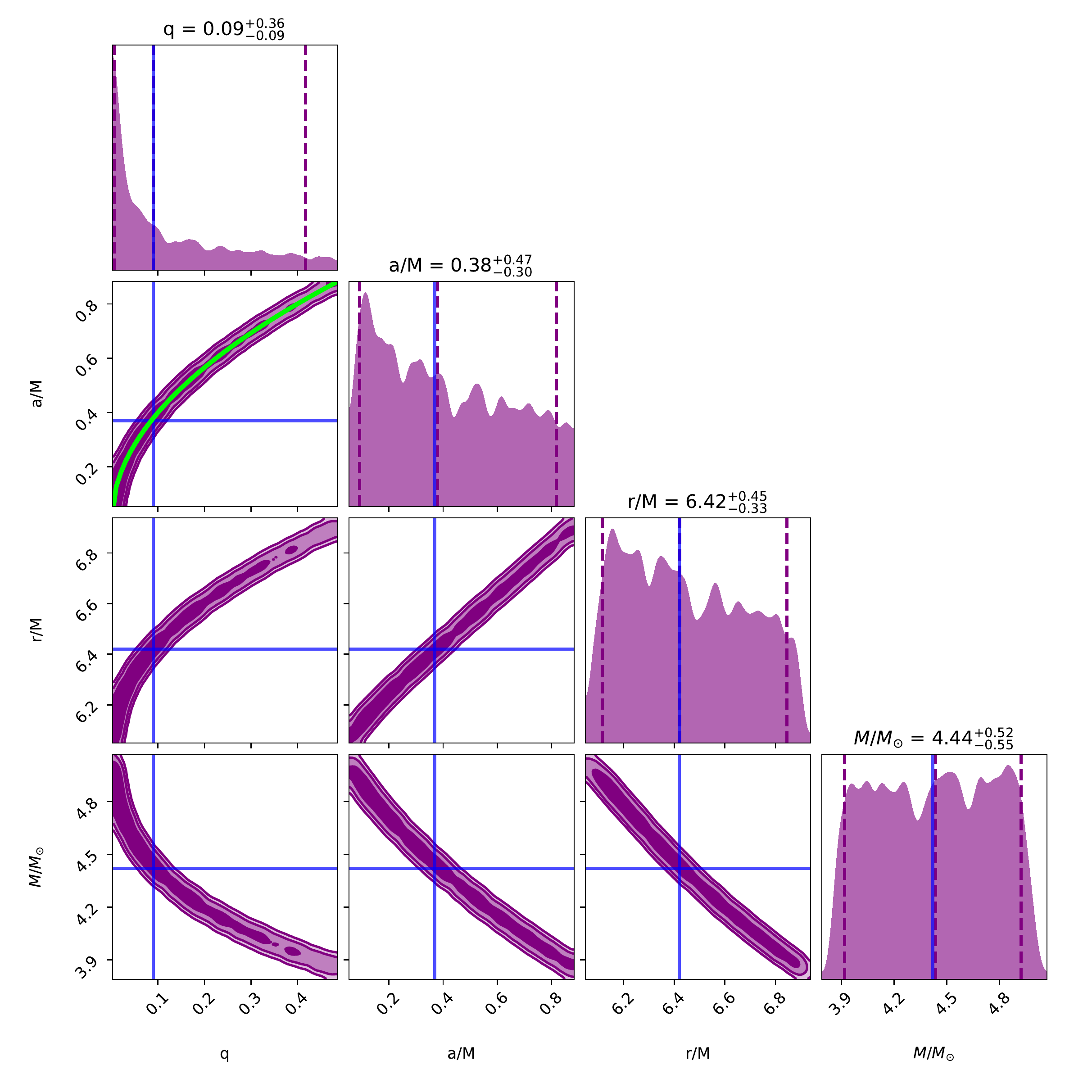}
		\caption{Same as figure.~\ref{figmc-2}, for model 1. Solid lines
		represent the mean values. The green curve represents the Kerr metric. The quadrupole agrees with the value  induced only by rotation. }
		\label{figmc-4}		
	\end{center}
	
\end{figure}
\begin{figure}[!ht]
	\begin{center}	
		\includegraphics[width=0.8\linewidth]{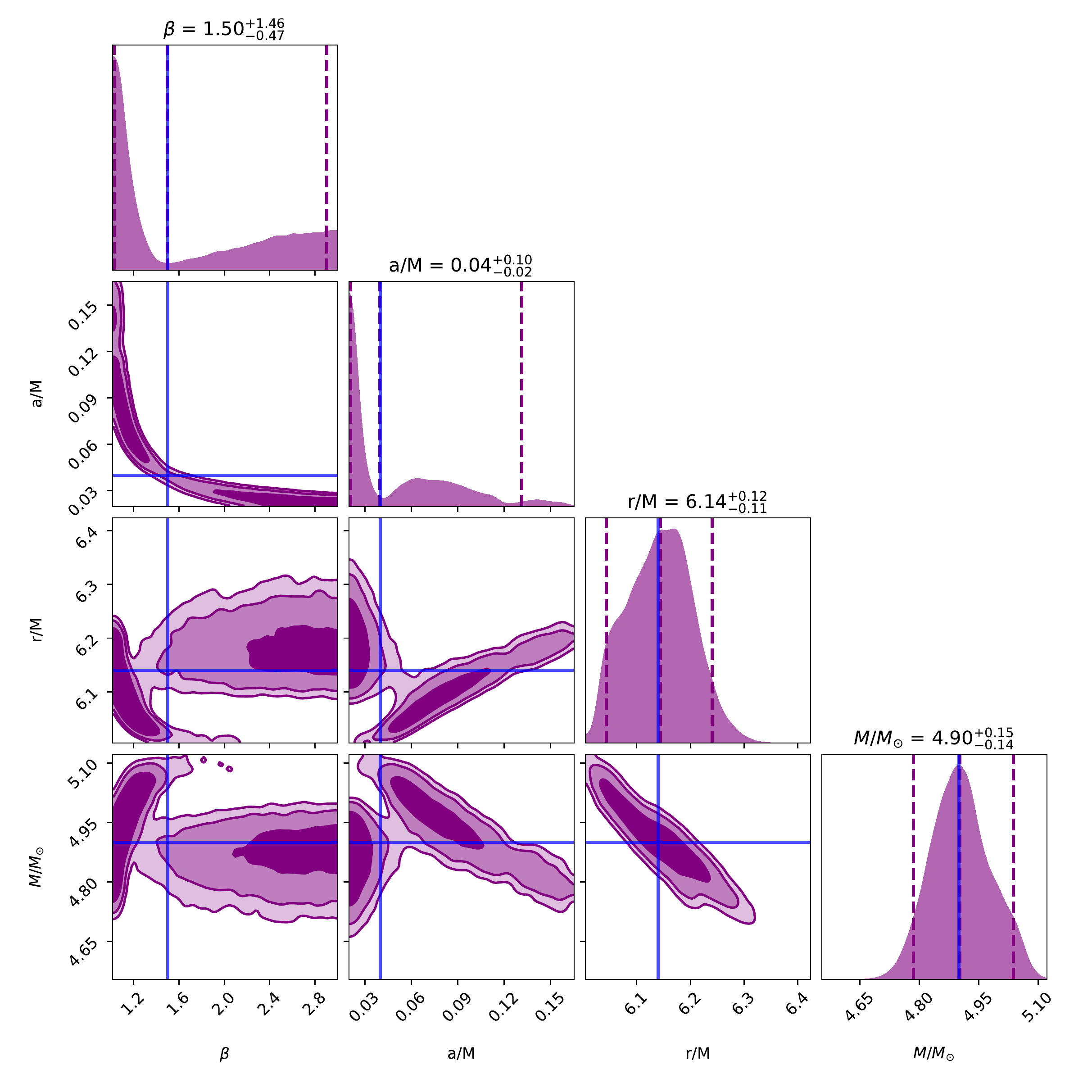}	
		\caption{Same as fig.~\ref{figmc-2}, for model 2. Solid lines
		represent the mean values.  }
		\label{figmc-6}			
	\end{center}	
\end{figure}

\subsection{Kerr-$Q$ metric}

Let us start with the Kerr-$Q$ metric.
Now let us introduce the posterior for the observation. We define the
posterior as
\begin{align}
\mathcal{P}\left(a/M,M, q,r|\nu_i \right)\propto \mathcal{L}\, p(q) \, p(a/M) \, p(M) \,p(r),
\end{align}
where we have non-informative flat priors for $0<q<0.9$ and $0<a/M<0.9$ and $4<r/M<10$. The mass for this system has been reported in  ref.~\cite{Beer:2001cg}. We use this measurement as our  prior and set $p(M/M_\odot)\sim\rm e^{-\frac{1}{2} \left(\frac{M/M_\odot-5.4}{0.3}\right) ^{2}}$ throughout this work \cite{Beer:2001cg}. We use \emph{dynesty}
for sampling the posterior \cite{dynesty}.

The covariance between variables is illustrated in figure.~\ref{figmc-2}. The
contours show 68\%, 95\% and 99\% credible intervals respectively. The posterior maximum for $q$ is the Kerr metric with $q=0$. The posterior mean values within 68\% credibility are $q=0.41^{+0.55}_{-0.39}$,
$a/M=0.4^{+0.08}_{-0.05}$ and $M/M_{\odot} =4.39^{+0.1}_{-0.15}$. Let us emphasize a crucial point here.  The reason that mass  is forced away from $5.4 \,M_\odot$  is because of truncation error and the complicated correlations among the parameters as we see in figure.~\ref{figmc-2}. This truncation  arises because of the expansion we have in $a/M$. The order of magnitude is $O\left((a/M)^{3} \right) $ and it introduces a truncation bias in the estimates. According to ref.~\cite{Motta:2013wga}, low-frequency QPOs, $\nu_{\rm n}$ here, for GRO
J1655$-$40 have varied over the years from $\sim 0.1$\,Hz reaching up to $\sim28 $\,Hz. At $r_{\rm ISCO}$ for the average values we find that $\nu_{\rm n}\sim23.97$\,Hz which is in the range observed.
According to the figure it is hard to discriminate the Kerr metric and
the Kerr-$Q$ metric. That is due to the specific combination of $q$ and $a/M$
that appears in the $\chi$-square, and there is a degeneracy in determining $q$
and $a/M$. In this regard, it is a combination of $a/M$ and $q$ that can
be measured more precisely.

\subsection{Hartle-Thorne metric}

In this section we construct two models based on the Hartle-Thorne metric. We use the same priors as in the previous section.
We also propose a novel test of no-hair theorem by directly constraining
the coefficient that appears in the metric as we will explain. Let us
consider two models based on the Hartle-Thorne metric in this section. In
the first model we treat $q$ as a free parameter and in the second model we
assume $q(a,M,\beta)={5\beta a^2}/{8M^2}$ where $\beta$ is related to
the inner structure of the compact object. If $\beta=1$ this is an evidence
that the object is a Kerr BH.

\subsubsection{Model 1}

As before we have the expression for $\chi$-square given in
eq.~\eqref{eq-chi20} and the observed values for the frequencies and the
errors are provided in eq.~\eqref{tab1}. The likelihood is also given by
$\cal L\sim\rm e^{-\chi^2/2}$.
The covariances are illustrated in figure.~\ref{figmc-4}. The mean of posterior values
are $q=0.09^{+0.36}_{-0.09}$, $a/M=0.38^{+0.47}_{-0.30}$ and $M/M_{\odot}
=4.44^{+0.52}_{-0.55}$. Here, we encounter the truncation error and  the complicated correlations among the parameters as we mentioned for the previous  model in our mass estimate. The green line shows the $q$ and $a/M$ for the Kerr metric. We note that the maximum of the posterior agrees with the Kerr metric. Also note that the posterior average obeys $q\sim (a/M)^2 $ which shows that all the quadrupole is induced by the rotation and is in agreement with the Kerr metric.
This time we have larger errors for the quadrupole. For Some values of the parameters within 1-$\sigma$ errors we have at $r_{\rm ISCO}$, $\nu_{n}\sim 27$\,Hz. Similarly we see that it is difficult to measure
$q$ with precision. It is a combination of $q$ and $a/M$ that can be
measured more precisely.

\subsubsection{Model 2}
In $model$-$1$, all the results on the green curve were consistent with the Kerr metric.
Here we propose a new test of no-hair theorem by adapting the Hartle-Thorne
metric. The model we consider here is a configuration in which
$q={5\beta a^2} / {8M^2}$, see ref.~\cite{Abbott:2020jks} for the importance of such  parameterization  in the context of gravitational waves. If $\beta=1$ we recover Kerr metric to second
order in $a$. This way we can compare the model with the Kerr model more directly. 
The covariances are illustrated in figure.~\ref{figmc-6}. The average of the posterior values
are $\beta=1.50^{+1.46}_{-0.47}$, $a/M=0.04^{+0.1}_{-0.02}$ and
$M/M_{\odot} =4.9^{+0.15}_{-0.14}$. Because we have $\beta a^2/M^2 $ in
the expression for $\chi$-square, we see that a combination of $\beta$ and
$a/M$ is confined with more precision. Please note the  truncation error and  the complicated correlations among the parameters as we mentioned for the previous  model in our mass estimate.  More
precise measurements or a combination of measurements could help decrease
the uncertainty in parameters. For a consistency check, note that the estimation of $q$ gives $q\sim 0.01$ which agrees within 68\% confidence with our result for $model$-$1$. The uncertainty is larger in this model. Using values of $a/M$ within $3$-$\sigma$ interval it is possible to get at $r_{\rm ISCO}$, $\nu_{n}\sim27$\,Hz.

\begin{table}[ht!]
	\begin{center}			
		\begin{tabular}{|c|c|
			}
			\hline
			$Model$&$\mathcal{R}$ 
			\\ \hline\hline			
			
			$Kerr$-$Q$&$0.702$ \\  \hline
			$Model$-$1$&$0.002$ \\ \hline
			$Model$-$2$&$0.014$\\ \hline
			
		\end{tabular}
		\caption{The Bayes factors for different models with respect to the Kerr model. }
		\label{Nu}
	\end{center}
\end{table}

\section{Model camparison}\label{m-c}

In the context of Bayesian inference, we are interested in the posterior $\mathcal{P}\left(\mathbf{\Theta}|\mathbf{D},\mathbf{M} \right) $ for a set of parameters $\mathbf{\Theta}$ for a given model $\mathbf{M}$ and data $\mathbf{D}$. We have
\begin{eqnarray}
\mathcal{P}\left(\mathbf{\Theta}|\mathbf{D},\mathbf{M} \right)=\frac{P(\mathbf{D}|\mathbf{\Theta},\mathbf{M})\mathbf{\pi}(\mathbf{\Theta}|\mathbf{M})}{P(\mathbf{D}|\mathbf{M})},
\end{eqnarray}
where $\mathbf{\pi}(\mathbf{\Theta})$ is the prior on $\mathbf{\Theta}$ and 
\begin{eqnarray}
P(\mathbf{D}|\mathbf{M})=\int P(\mathbf{D}|\mathbf{\Theta},\mathbf{M})\mathbf{\pi}(\mathbf{\Theta})d^{}\mathbf{\Theta}.
\end{eqnarray}
Similarly for a model we have $\mathcal{P}(\mathbf{M}|D)\propto P(\mathbf{D}|\mathbf{M})\mathbf{\pi}(\mathbf{M})\propto \mathcal{Z}_{\mathbf{M}}\mathbf{\pi}_{\mathbf{M}} $ where
$\mathcal{Z}_{\mathbf{M}}=P(\mathbf{D}|\mathbf{M})$ and $\mathbf{\pi}_{\mathbf{M}}=\mathbf{\pi}(\mathbf{M})$. $\mathcal{Z}_{\mathbf{M}}$ is the model evidence and $\mathbf{\pi}_{\mathbf{M}}$ is the model prior.  To compare the  models, we define the Bayes factor as
\begin{eqnarray}
 \mathcal{R}=\frac{\mathcal{Z}_{\mathbf{M_1}}\mathbf{\pi}_{\mathbf{M_1}}}{\mathcal{Z}_{\mathbf{M_2}}\mathbf{\pi}_{\mathbf{M_2}}}.  
\end{eqnarray}
If $\mathcal{R}>1$, there is evidence that $\mathbf{M_1}$ is better than $\mathbf{M_2}$ in describing data.
Here we select the Kerr metric for our comparison and consider flat priors for all the models. The Bayes factor for each model is reported in Table~\ref{Nu}.
For Kerr-$Q$ metric we have $\mathcal{R}_{\rm Kerr-Q}=0.7$. We find that the Kerr metric is preferred over the Kerr-$Q$  metric. For $model$-$1$, we have $\mathcal{R}_{1}=0.002$. 
For $model$-$2$, we have $\mathcal{R}_{2}=0.014$.
Although there is very strong evidence that Kerr-$Q$ metric is better than $model$-$1$ and $model$-$2$
, we find that there is generally strong evidence for the Kerr metric as the source for describing the QPOs in GRO J1655$-$40.

\section{Summary and discussion}\label{s-d}

We have provided models to test the no-hair theorem by considering three
families of solutions based on two metrics that describe a rotating object
augmented with a quadrupole. The first case study and the first metric is
the Kerr-$Q$ metric derived from approximating an exact solution. We
encounter a degeneracy to constrain the spin and the quadrupole. This
hinders our ability to measure the quadrupole with more precision. This
makes it hard to discriminate the Kerr or beyond-Kerr nature of the compact
object. As a result we find that the Kerr BH lies in the 
credible region in the allowed parameter space. However, $\mathcal{R}_{\rm Kerr-Q}=0.7$ with respect to Kerr. The second metric is the
Hartle-Thorne solution. This metric describes the exterior of rotating
compact objects with some quadrupole. As our second study we treat the
quaropole as a free parameter. Again, we find a degeneracy in constraining
the spin and the quadrupole and consistency with a Kerr BH. In this case we find $\mathcal{R}_{1}=0.002$. This way, there is no evidence that this model is better than the Kerr model for the central object.

In our third case study, we devise a novel test of no-hair theorem by
treating the quadrupole as a function of the spin by introducing a
parameter $\beta$. If the object is a Kerr BH, then $\beta=1$. We find that
although best fit weakly hints towards some other compact object rather
than the Kerr BH, the Kerr BH lies well in the 3-$\sigma$ credible interval in
the parameter space. Here we find that  $\mathcal{R}_{2}=0.02$. Although this model performs better than $model-1$, there is strong evidence that the Kerr model is true. The bottom line is that we found strong evidence in favor of the Kerr metric. Thus, there is strong evidence  that no-hair theorem holds for GRO J1655$-$40.

 We used test mass orbits in the relativistic precession model. In reality, there are complications in modeling the accreting  material that could affect QPOs \cite{Ingram:2020aki}.  Further theoretical and numerical frameworks need to elucidate the environmental effects such as thermal effects and optical thickness of  accreting  material on the QPOs \cite{Ingram:2014ara}.  Such effects could hinder attempts to address an observed discrepancy with the Kerr metric to some generalized Kerr metric. Without complementing studies, the gas emission around  a non-Kerr object with some spin could be similar to a Kerr object with some other spin and a combination of different methods may help. Thus, other studies such as study of the disk’s thermal spectrum and the broad K$\alpha$ iron line might help to shed light on the nature of the compact object \cite{Bambi:2013fea}.
 
\section*{Acknowledgments}

We thank the anonymous referee for constructive comments. This work was supported by the National Natural Science Foundation of China
(11991053, 11721303, 11975027), the Young Elite Scientists Sponsorship
Program by the China Association for Science and Technology (2018QNRC001),
the National SKA Program of China (2020SKA0120300), and the Max Planck
Partner Group Program funded by the Max Planck Society.

\end{document}